# Solar-cycle and Latitude Variations in the Internetwork Magnetism

J. C. Trelles Arjona[1], M. J. Martínez González[2,3], and B. Ruiz Cobo[2,3]
[1] Laboratory for Atmospheric and Space Physics (LASP), University of Colorado, 1234 Innovation Dr, Boulder, CO 80303, USA; Juan.Trelles@lasp.colorado.edu
[2] Instituto de Astrofísica de Canarias (IAC), Vía Láctea s/n, E-38205 San Cristóbal de La Laguna, Tenerife, Spain; marian@iac.es, brc@iac.es
[3] Dept. Astrofísica, Universidad de La Laguna, E-38205 San Cristóbal de La Laguna, Tenerife, Spain

## Abstract

The importance of the quiet-Sun magnetism is that it is always there to a greater or lesser extent, being a constant provider of energy, independently of the solar cycle phase. The open questions about the quiet-Sun magnetism include those related to its origin. Most people claim that the local dynamo action is the mechanism that causes it. This fact would imply that the quiet-Sun magnetism is nearly the same at any location over the solar surface and at any time. Many works claim that the quiet Sun does not have any variation at all, although a few of them raise doubt on this claim and find mild evidence of a cyclic variation in the the quiet-Sun magnetism. In this work, we detect clear variations in the internetwork magnetism both with latitude and solar cycle. In terms of latitude, we find an increase in the averaged magnetic fields toward the solar poles. We also find long-term variations in the averaged magnetic field at the disk center and solar poles, and both variations are almost anticorrelated. These findings do not support the idea that the local dynamo action is the unique factory of the quiet-Sun magnetism.

*Unified Astronomy Thesaurus concepts:* Quiet sun (1322); Solar cycle (1487); Solar dynamo (2001); Solar magnetic fields (1503); Solar physics (1476)

## 1. Introduction

Active regions (hereinafter ARs) are easily recognized over the solar photosphere. The strong magnetic fields contained in it deform the granulation and eventually concentrate as dark, cold sunspots. Far from these areas, the solar photosphere is known as the quiet Sun, and it is characterized by serene granulation. This denomination comes from the very first magnetographs, whose lack of spatial resolution (and polarimetric sensitivity) did not allow detecting the weakest magnetic fields of the solar atmosphere. The development of more sensitive instrumentation allowed the discovery of a quiet Sun that is fully magnetized (Bellot Rubio & Orozco Suárez 2019).

The quiet Sun is composed of the network and the internetwork (hereinafter IN). The former contains the strongest fields in the form of thin vertical tubes that are located at the borders of supergranular cells. The latter is a collection of weak, small-scale, disorganized fields over the solar surface, and their empirical study is therefore a great challenge. Still, it is crucial to deepen the knowledge of IN magnetic fields because they populate almost the entire solar surface at any time during the solar cycle, having an important contribution to the magnetic energy budget (Trujillo Bueno et al. 2004) and likely having implications for the heating of the chromosphere (Gömöry et al. 2010; Martínez González et al. 2010). This article focuses on two subjects that have been treated very little in the literature and will allow us to unveil the origin and nature of IN fields: the variation in the IN magnetism with solar cycle and with solar latitude.

The activity of the solar magnetism generated by the interior dynamo, such as the global field and ARs, varies with an 11 yr period (22 for the polarity of the field). The ARs are clearly bounded to the global dynamo. However, some studies claim that IN magnetic fields are created by an independent dynamo that is working at surface layers, i.e., the so-called surface dynamo. They are therefore not subjected to the 11 yr cycle. Sánchez Almeida (2003) compared two spectropolarimetric data sets taken near the disk center at maximum (2003) and minimum (1996) periods and found no significant variations in the IN fields between them. Buehler et al. (2013) used data taken regularly by the Solar Optical Telescope on board the Hinode satellite (SOT; Kosugi et al. 2007; Ichimoto et al. 2008; Shimizu et al. 2008; Suematsu et al. 2008; Tsuneta et al. 2008) during 6 yr (2006 to 2012) near the disk center, and they found no significant differences.

In contrast to these results, there are works that claimed the detection of variations in the IN magnetic fields with solar cycle. Kleint et al. (2010) used observations in the spectral region around 5141 Å to infer the mean strength of small-scale magnetic fields via the differential Hanle effect in molecules (Berdyugina & Fluri 2004). They conducted a two-year synoptic program starting in 2007 December (solar minimum), and they did not find variations in the strength of the small-scale magnetic fields. But by comparing with observations taken by other authors at the solar maximum, they found that the averaged strength of IN magnetic fields may have mildly weakened from solar maximum to solar minimum. Faurobert & Ricort (2015) performed a Fourier spectral analysis to find differences in the spatial structures in the maps of the unsigned circular and linear polarization of IN magnetic fields between observations taken in 2007 (solar minimum) and 2013 (solar maximum) with the SOT on board the Hinode satellite. They detected a marginally significant difference between the power spectra in 2013 and 2007 (the fluctuation was weaker in 2013). Thus, a firm detection of the variation in the quiet-Sun magnetism with time has not been reported so far. In this letter, we present the first clear detection of these variations.

Concerning the latitude variation of the quiet-Sun magnetism, Martínez González et al. (2008) used data taken at several positions over the solar surface with the Tenerife Infrared

**Table 1**
Observational Details of the Data Sets Used in This Work

| ID | Date | Time (UT) | Heliographic Coordinates (x; y) (arcsec) | Solar Latitude (degrees) | FoV (arcsec) | Exposure Time (s) | ⟨**B**⟩ (G) |
|---|---|---|---|---|---|---|---|
| DC1 | 2018 Aug 29 | 8:25 | 0″; 0″ | 7.14 | 58.″0 × 54.″0 | 2.0 | 77.45 |
| NP1 | 2018 Sep 1 | 8:08 | 0″; 940″ | 68.71 | 58.″0 × 54.″0 | 2.0 | 141.07 |
| SP1 | 2018 Sep 1 | 8:47 | 0″; −900″ | −55.51 | 58.″0 × 54.″0 | 2.0 | 129.11 |
| DC2 | 2018 Dec 14 | 12:09 | 0″; 0″ | −0.81 | 58.″0 × 54.″0 | 1.2 | 98.74 |
| MU181 | 2018 Dec 14 | 10:48 | 0″; 400″ | 23.39 | 58.″0 × 40.″5 | 1.2 | 97.25 |
| MU182 | 2018 Dec 14 | 10:29 | 0″; 600″ | 37.13 | 58.″0 × 40.″5 | 1.2 | 59.90 |
| MU183 | 2018 Dec 14 | 10:11 | 0″; 860″ | 60.99 | 58.″0 × 37.″8 | 1.2 | 41.22 |
| NP2 | 2018 Dec 14 | 11:35 | 0″; 930″ | 71.62 | 58.″0 × 40.″5 | 1.2 | 63.03 |
| SP2 | 2018 Dec 17 | 11:12 | 0″; −950″ | −77.13 | 58.″0 × 47.″0 | 1.2 | 108.50 |
| NP3 | 2018 Dec 17 | 9:35 | 0″; 950″ | 76.86 | 58.″0 × 54.″0 | 1.2 | 68.11 |
| DC3 | 2019 Sep 19 | 7:45 | 0″; 0″ | 7.14 | 30.″0 × 30.″0 | 1.2 | 116.71 |
| SP3 | 2019 Sep 19 | 8:11 | 72″; −927″ | −57.43 | 30.″0 × 30.″0 | 1.2 | 92.95 |
| NP4 | 2019 Sep 19 | 8:24 | 71″; 919″ | 64.24 | 30.″0 × 30.″0 | 1.2 | 102.65 |
| DC4 | 2020 Nov 19 | 8:41 | 0″; 0″ | 2.30 | 58.″0 × 27.″0 | 1.2 | 138.84 |
| MU201 | 2020 Nov 19 | 10:35 | 0″; 600″ | 38.19 | 58.″0 × 27.″0 | 1.2 | 104.22 |
| MU202 | 2020 Nov 19 | 10:19 | 0″; 800″ | 53.58 | 58.″0 × 27.″0 | 1.2 | 91.36 |
| NP5 | 2020 Nov 19 | 9:51 | 0″; 940″ | 68.06 | 58.″0 × 54.″0 | 1.2 | 193.56 |
| DC5 | 2021 Aug 16 | 8:45 | 0″; 0″ | 6.70 | 30.″0 × 48.″0 | 1.2 | 127.55 |
| NP6 | 2021 Aug 16 | 11:20 | 0″; 920″ | 71.08 | 30.″0 × 30.″0 | 1.2 | 164.85 |
| SP4 | 2021 Aug 16 | 11:50 | 0″; −920″ | −61.55 | 30.″0 × 30.″0 | 1.2 | 145.07 |
| EL1 | 2021 Aug 16 | 11:07 | 920″; 0″ | 17.67 | 30.″0 × 30.″0 | 1.2 | 126.72 |

**Note.** From left to right: Data set identification, observation date, starting time of the scan, heliocentric coordinates and solar latitude of the center of the scan, FoV, exposure time, and averaged magnetic field strength.

Polarimeter (Martínez Pillet et al. 1999) at the Vacuum Tower Telescope. At their spatial resolution (∼1″), no significant variations in the properties of IN magnetism on the data sets were observed at different heliocentric angles. Ito et al. (2010) used data from SOT to compare two observations at the north pole and the east limb. These two positions allowed them to remove projection effects from the analysis and retain the latitude variation because both sets have the same heliocentric angle. They found that the total magnetic flux of the east limb was lower than that of the north polar region.

The most complete work until now concerning latitude-time variations in the IN magnetism is the work of Lites et al. (2014). They used data taken from SOT to study the IN magnetism variations during the period 2008–2013. They did not find variations with time at moderate latitudes. However, they found changes in the polarity imbalance and its time variation near the poles. Moreover, they found an increase in both longitudinal and transversal magnetic flux in the weakest component of IN magnetism toward the poles.

In this work, we present a complete study both in latitude and cycle variations of the IN magnetic fields using the infrared spectral lines at 1.5 $\mu$m. This spectral window traces the deep photosphere with the best magnetic field sensitivity (spectral line at 15648.514 Å with a Landé factor of 3). Here, we use observations taken along 4 yr at several heliocentric angles to shed some light on the open questions in this essential branch of solar magnetism.

## 2. Observations

The set of observations presented in this work (see Table 1) was carried out during four consecutive years (2018, 2019, 2020, and 2021). Thus, the time span between the first and last observation is 3 yr. We used the GRIS (Collados et al. 2007, 2012) spectropolarimeter, installed at the German GREGOR telescope (Schmidt et al. 2012), to record the four Stokes parameters in the spectral range around 1.565 $\mu$m. In all observing campaigns, we used the same spectral sampling (40 mÅ), spectral range (30 Å, from 15644 to 15674 Å), and exposure time (30 ms, except for the data taken in the first campaign of 2018, where the exposure time was 50 ms). The only difference between the observing campaigns was the use of either the slit (in 2018 and 2020) or the Integral Field Unit (IFU; in 2019 and 2021). The slit or IFU position was aligned with the solar north–south direction. The adaptive optics system (Berkefeld et al. 2016) locked on granulation allowed a spatial resolution of about 0.″5. The field of view (FoV) of the slit and the IFU observations are 0.″135 × 60″ and 3″ × 6″, respectively.

We used the dedicated software from Schlichenmaier & Collados (2002) to demodulate the data. The same software allowed us to correct for bias, flatfield, and bad pixels and to remove the instrumental crosstalk. We applied additional corrections to the data: subtraction of the wavelength-independent stray-light contamination from the intensity spectra, removal of the residual crosstalk from Stokes I from Stokes Q, U, and V, reduction of noise from the data using a principal component analysis (Loève 1955; Rees & Guo 2003), and correction of polarized interference fringes (see Trelles Arjona et al. 2021b, and references therein for more details of the reduction process).

In Figure 1 we show an example of the observed data sets. In particular, we show DC2, MU182, and NP2 (see Table 1 for details), corresponding to the observations at solar latitudes $\Lambda = -0°.81, 37°.13,$ and $71°.62$ (where the negative sign means toward the south pole), respectively. As can be seen, most of the FoV has polarimetric signals, both in linear and circular polarization, even close to the limb.

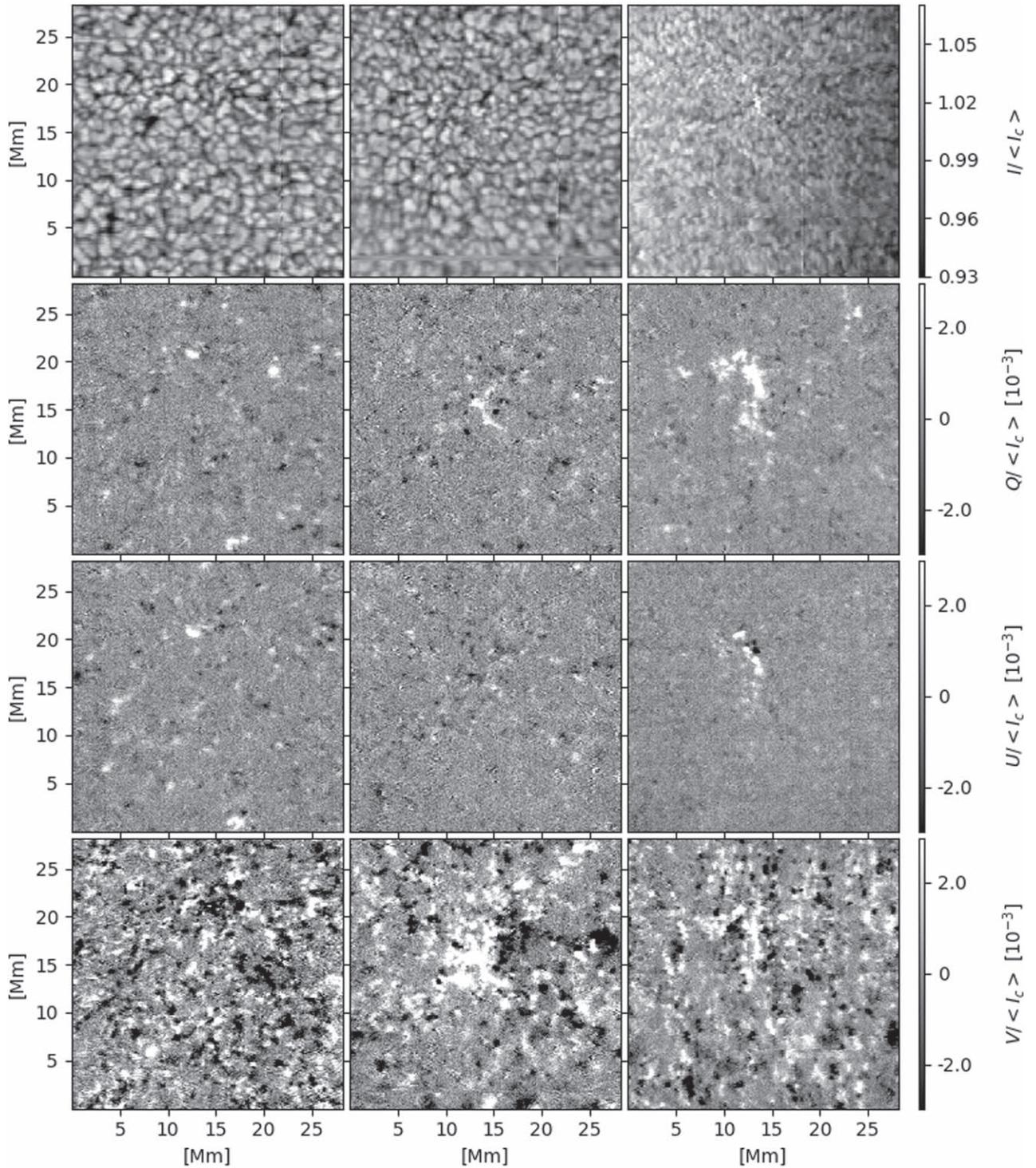

**Figure 1.** Example of observations DC2, MU182, and NP2 (left to right). From top to bottom, we display the intensity and Stokes *Q*, *U*, and *V* amplitudes.

## 3. Inference of Physical Parameters from Data Inversions

In our case, we have used the code called Stokes inversion based on the response functions (SIR; Ruiz Cobo & del Toro Iniesta 1992) to infer the stratification of the physical quantities of the solar atmosphere. SIR looks for the synthetic profile that better matches a given observation assuming local Tthermodynamic Eequilibrium, which is a very suitable approximation for spectral lines that formed in the low photosphere, such as the lines at 1.5 $\mu$m. SIR synthesizes the profiles using a parameterized model atmosphere and solving the radiative transfer equation. The initial model atmosphere is modified by SIR to minimize the difference between the synthesis and the observation.

The inversion strategy in this work is based on the simultaneous inversion of the 15 spectral lines in our spectral range (see Table 1 in Trelles Arjona et al. 2021b) using only the intensity profiles. The strategy assumes one magnetic atmosphere occupying a fraction of the resolution element (23%) and stray light filling the rest of space (77%; see more details of the stray light treatment in Trelles Arjona et al. 2021a). The free atmospheric parameters in the inversion are the temperature, line-of-sight (LoS), and microtubulent velocities, and the magnetic field strength and inclination. We allow variations with height in all parameters. SIR can select, through the number of nodes, the complexity of these gradients for each pixel by considering the information of the observed Stokes profiles and the response functions (see more details in del Toro Iniesta & Ruiz Cobo 2016). It has been demonstrated in Trelles Arjona et al. (2021a) that this strategy is very suitable to infer the strength of quiet-Sun fields where the magnetic fields are still not completely resolved (due to the lack of spatial resolution in current instrumentation). We selected as initial atmosphere for the inversion model C from Fontenla et al. (1993; FALC model). In each inversion, the initial model atmosphere was randomly changed in LoS and microtubulent velocities, (between the interval 0 and 5 km s$^{-1}$) and in magnetic field strength and inclination (between 0 and 1500 G, and 0° and 180°). We ran 45 inversions per pixel, and the inversion with the lowest $\chi^2$ was proposed as the solution.

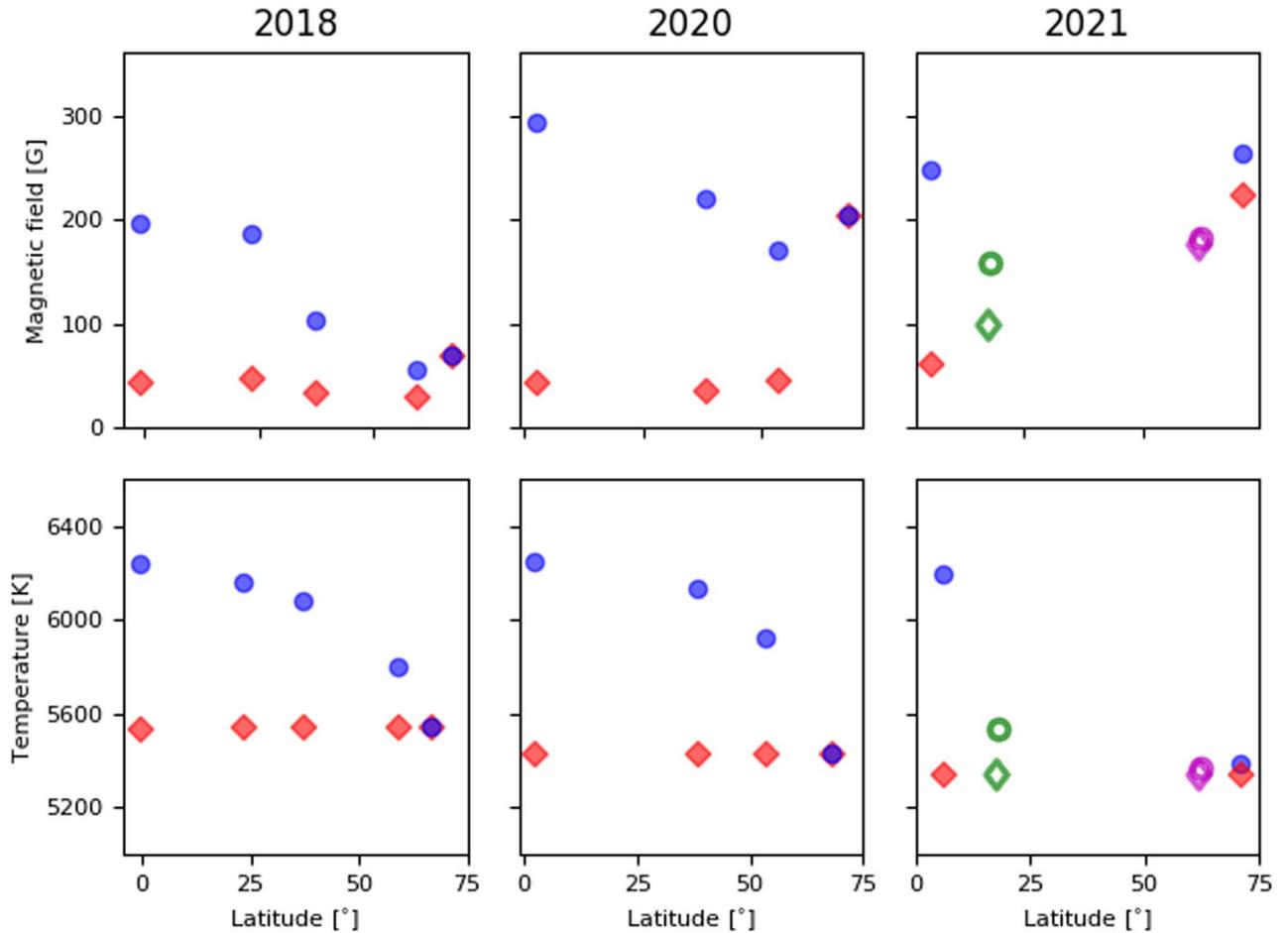

**Figure 2.** Mean magnetic field and temperature variations with latitude in 2018, 2020, and 2021. Top row: mean magnetic field in Gauss. Bottom row: mean temperature in Kelvin. Circles represent averaged values inferred in the inversion without correction for optical depth. Diamonds depict averaged values inferred in the inversion corrected in optical depth after introducing an optical depth shift to probe the same atmospheric layer. In green we display the results of a data set in the east limb (i.e., low $\mu$ and low latitude). In magenta we present the results of a data set at the south solar pole (i.e., in absolute value of latitude for the sake of clarity).

## 4. Results

The average magnetic field strength inferred from all the data sets is listened in the right column of Table 1. These numbers are obtained by averaging the full FoV and the optical depth $0.0 \geqslant \log(\tau) \geqslant -1.2$, which is the sensitivity range of the spectral lines (at the disk center) as deduced from response functions (Trelles Arjona et al. 2021b).

### 4.1. Latitude Variation of the Quiet-Sun Magnetism

The heliocentric angle ($\theta$; the angle between the LoS and the normal to the solar surface) is usually given in terms of $\mu = \cos\theta$. As the inclination of the LoS with respect to the solar vertical increases (i.e., $\theta$ increases and $\mu$ decreases), photons reach optical depth unity higher in the atmosphere. Hence, to compare similar layers of the solar atmosphere in observations with different heliocentric angles, a shift in the optical depth axis at 5000 Å is needed. In particular, one needs to remove an offset given by $\log_{10}(\mu)$. In this way, for example, to compare the inversion results of an observation with an heliocentric angle $\mu = 0.25$ with results of inversions performed at disk center observation, we take the mean value of solar atmospheric parameters from 0.0 to $-0.4$ and $-0.6$ to $-1.0$ in $\log(\tau)$ units, respectively.

The variation in the average magnetic field strength and temperature with solar latitude is displayed in Figure 2. We have displayed the results of observations at different latitudes that were observed on the same day. We have displayed the data sets obtained in 2018 December 14, 2020 November 19, and 2021 August 16 because on those days, the latitudes were covered well. Red diamonds and blue circles represent the inversion results with and without correction for optical depth, respectively. Green circles and diamonds that can be seen in the right column plots represent the results obtained from a data set observed at the east limb of the Sun. The comparison with limb observations at equatorial latitudes serves to discard geometrical effects when analyzing polar data. Magenta circles and diamonds represent the results from the south pole (we use the absolute value of the latitude in this case for visualization purposes).

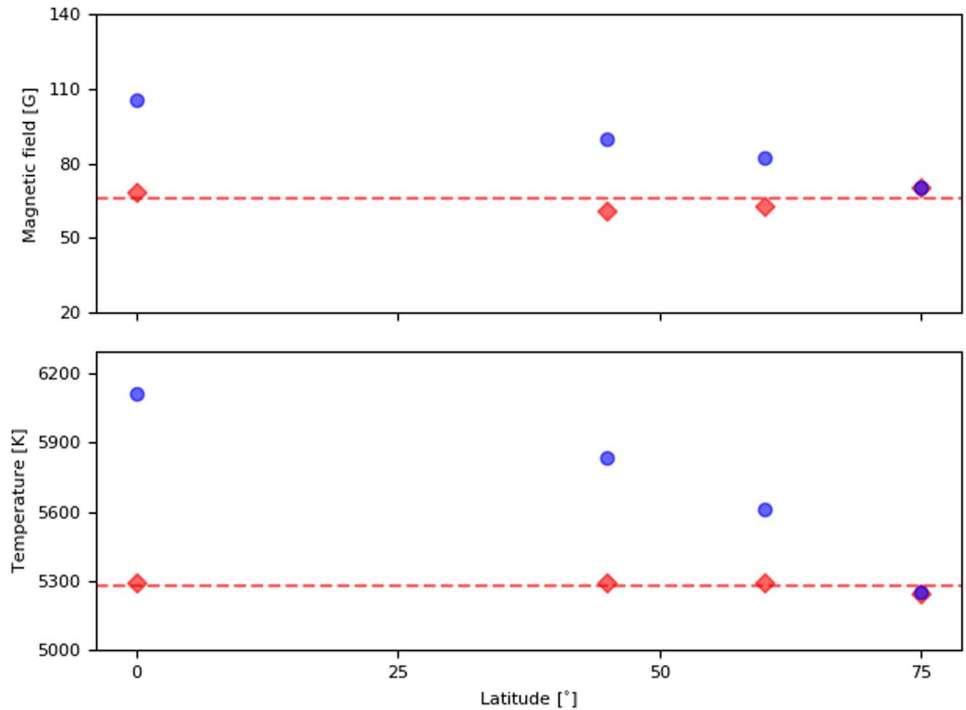

**Figure 3.** Mean values retrieved from inversions of profiles synthesized from a numerical simulation for the magnetic field strength (upper panel) and temperature (lower panel) at different latitudes. Blue circles represent averaged values at the same optical depth ($0.0 \geqslant \log(\tau) \geqslant -0.4$). Red diamonds stand for averaged values corrected for optical depth to observe the same layers at different simulated latitudes.

If we look at the circles, i.e., at the averages obtained at the same $\log \tau$ range, a clear decrease in temperature with latitude is observed. This means that the temperature is lower as we reach higher atmospheric layers, which is the expected behavior in the solar photosphere. The average magnetic field follows a similar trend up to the poles. We indeed expect a dilution of the magnetic field with height due to flux conservation and field expansion. However, when we reach the poles, we detect an increase in the average field. This occurs to a greater or lesser extent on the three days of observation. Similar results were obtained in Lites et al. (2014). They calculated the unsigned longitudinal apparent flux density (see their Figure 4(d)) and the transverse apparent flux density (see their Figure 7(c)). In both cases, an increase in apparent flux is observed at the poles.

Looking at the diamonds that represent the results at a similar height (shifted LoS optical axis), we see that as expected, the temperature is very similar at all latitudes. The values of the magnetic field strength do not show significant variations either, until, again, we reach the poles. Interestingly, the results from the east limb (polar geometry, but the same latitude as the disk center observations) show values similar to those of the disk center data, except for the poles, which show an increase in field strength at the north and south poles.

In order to check the reliability of the increase in the average magnetic field at the poles, we have synthesized and inverted the target spectral lines using the SIR code (SIR; Ruiz Cobo & del Toro Iniesta 1992) in a MANCHARAY magnetohydrodynamical quiet-Sun simulation (see details of the numerical set up of the simulation in Khomenko et al. 2017). We have degraded the synthetic data to the spatial resolution of the observations and introduced random noise at the level of the observational one. We have inclined the simulation cube to mimic observations at different heliocentric angles. We use four heliocentric angles $\mu = 1.00$, 0.70, 0.50, and 0.25 (the corresponding latitudes are $\Lambda = 0°$, 45°, 60°, and 75°, respectively). We took the pixel size increase due to the geometrical effects into account. By way of example, at disk center, the sampling required to mimic GRIS observations in the simulation is $5 \times 5$ pixels, while at $\mu = 0.5$, the sampling is $10 \times 5$ pixels. Even though the atmospheric gradients in the simulations are very complex, the inversion with the SIR code captures the trend of the gradients.

The inferred values of the average magnetic field strength and temperature from synthetic spectra are shown in Figure 3. As before, blue circles stand for the inversion results without correction (i.e., results averaged at the same optical depth range $0.0 \geqslant \log(\tau) \geqslant -0.4$), thus, tracing higher layers of the atmosphere as the latitude increases. Both temperature and average magnetic field decrease with height, as expected for physical reasons and as it is the case of the real gradients of the simulations. No increase in the magnetic field strength average is detected at high inclinations, which corroborates the validity of our inversion procedure. Red diamonds represent the inversion results shifted in optical depth to trace a similar geometrical layer of the atmosphere. We infer an almost constant value of the average magnetic field strength and temperature (dashed red line), which is the expected behavior.

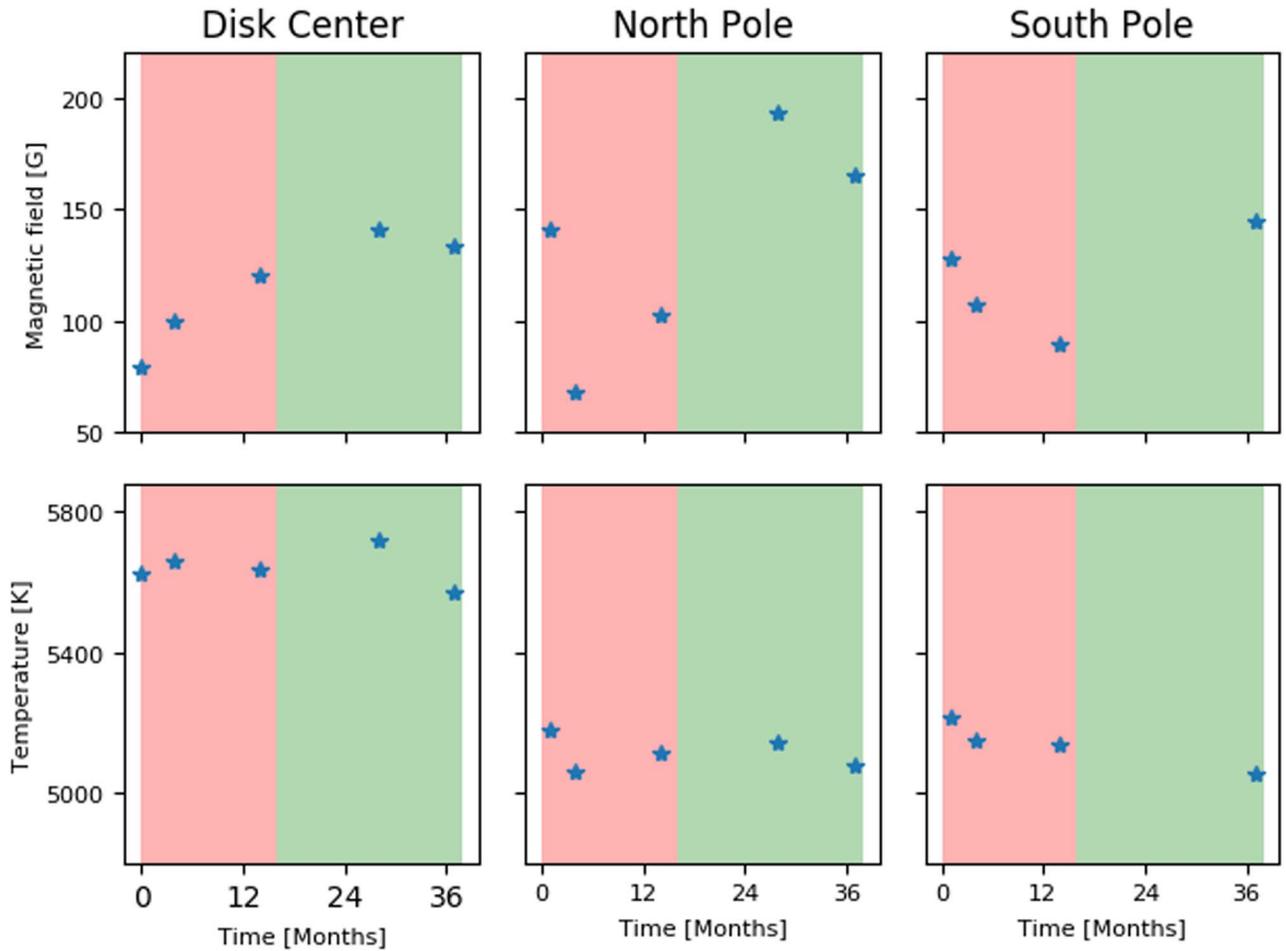

**Figure 4.** Mean magnetic field and temperature variations between 2018 August and 2021 August. Top row: mean magnetic field in Gauss. Bottom row: mean temperature in Kelvin. From left to right: Disk center and the north and south poles. Red and green stand for the decreasing and increasing phase of the global magnetic cycle of the Sun, respectively.

Again, we do not detect an increase of the average field strength at the near-limb geometries. This makes us confident that the sudden increase in the field at the poles detected in observations is of solar origin.

### 4.2. Time Variation in the Quiet-Sun Magnetism

The time evolution of the averaged magnetic field strength and temperature at the disk center and the poles is displayed in Figure 4. The initial time is 2018 August 29. The decaying and rising cycle phase are represented in pink and green color, respectively. The boundary between these two phases corresponds to the minimum of activity at solar cycle 25 (2019 December), which was estimated with the smoothed monthly values of the sunspot number (SILSO data/image, Royal Observatory of Belgium, Brussels).

The average temperature remains almost constant over time both at disk center and at the poles, i.e., there are no evident variations in the average photospheric temperature with solar cycle. The values of the temperature at disk center are higher than those at the poles. This means that the temperature is lower at higher atmospheric layers, which is what we expect at the solar photosphere. The scenario is different if we consider the average magnetic field strength. At the disk center, we observe a rising trend in the magnetic field strength from 2018 to 2021. At the poles, the average field diminishes in the decaying phase and increases in the rising phase. The behavior at the poles is in phase with the activity cycle, while at disk center, the rising trend of the average magnetic field starts at least 1 yr prior to the activity minimum.

### 5. Conclusions and Discussion

The high quality of GRIS data together with our analysis procedure has allowed the precise determination of the average magnetic field of the quietest areas of the solar surface. By analyzing data taken at different latitudes and different times, we claim a clear detection of a cycle and latitude variation in the strength of the quiet-Sun magnetism. The quiet-Sun magnetism is highly variable in the range between 40 and 200 G, with a mean value of 109 G. It depends on both time and location over the solar surface.

Previous claims of latitude variation of the quiet-Sun magnetism were performed with the longitudinal and transverse magnetic field (Lites et al. 2014). Lites et al. find that the transverse magnetic flux density (see their Figure 7(c)) decreases with latitude, except for the poles, where it increases substantially. Here, we confirm that also the strength of the magnetic fields of the quiet Sun follows the same trend. The decrease in the average field strength with latitude is consistent with the expansion of the fields with height, but the sudden increase at the poles is still a matter for further research.

The cycle variation in the quiet-Sun magnetism has been a long-standing unknown, with many works showing opposing results, and the few detections were not very highly significant (Kleint et al. 2010; Jin et al. 2011; Faurobert & Ricort 2015). In this article, we present the first clear detection of a time variation in the IN magnetism. It may be a very reasonable assumption that the main driver behind these time variations is the solar activity cycle. Moreover, this variation depends on solar latitude. At the poles, the variation is in phase with the activity cycle, i.e., the field strength decreases toward the activity minimum. However, at disk center, the averaged magnetic field starts to increase before the global activity cycle starts to rise.

Further work is necessary for a more robust interpretation of our findings. It would be advisable to complete the whole solar cycle with the same strategy of the observations and inversions, for example, to confirm whether the last measurement of the average magnetic field in the disk center indeed shows a decaying phase. However, based on the results obtained so far, we propose a possible scenario in which the quiet-Sun magnetism has its origin in the global dynamo.

Prior to the beginning of a new cycle, magnetic flux is transported from the interior to the surface. At this time, the only way to detect it is using helioseismology, as pointed out by Korpi-Lagg et al. (2022). In this work, they predict the flux to be emerging at disk center in mid-2017, which is one year prior to the minimum field strength that we inferred in our data series at disk center. This new flux appears at the surface as mixed polarity fields because turbulent granular motions in the upper convection zone disrupts the emerging tubes quite efficiently (Cheung et al. 2007). We have detected that as the cycle evolves toward the sunspot minimum, the average field strength at the surface increases. We therefore speculate that the field strength of the emerging flux tubes also increases, allowing the formation of large structures such as sunspots (Cheung et al. 2010).

We also speculate that the meridional flow (a circular flow of solar plasma from the equator to the poles in the upper convection zone and backward in the bottom of the convection zone, close to the tachocline) can transport the decaying flux from the following sunspots to the poles. This means that the expected behavior of the average magnetic field at the poles will follow the sunspot cycle, which is the case in our data. This may seem to contradict previous results that stated that the dipolar field is maximum at sunspot minima. However, it is just an apparent contradiction because our average field strength is not subject to cancellations, while it is the case for the longitudinal flux on which previous works relied. Along the sunspot cycle, the opposite polarity to that at the poles is transported by the meridional flux. When observing circular polarization at moderate spatial resolution as compared to the scales of these quiet-Sun fields (a few kilometers), we are blind to most of the fields because of cancellations. As more and more fields are dragged to the poles, the polarity of these fields start to dominate until the polarity of the pole is reversed. With our procedure, we are sensitive to all the fields within the resolution element, which is the main reason that we see the correlation of the quiet-Sun fields at the poles with sunspot cycle.

The authors are especially grateful to Prof. Manuel Collados Vera for very interesting discussions and comments to improve the manuscript. We acknowledge financial support from the Spanish Ministerio de Ciencia, Innovación y Universidades through project PGC2018-102108-B-I00 and FEDER funds. J.C.T.A. acknowledges financial support by the Instituto de Astrofísica de Canarias through Astrofísicos Residentes fellowship and the University of Colorado Boulder through the George Ellery Hale fellowship. M.J.M.G. acknowledges financial support through the Ramón y Cajal fellowship. The observations used in this study were taken with the GREGOR telescope, located at Teide observatory (Spain). The 1.5 m GREGOR solar telescope was built by a German consortium under the leadership of the Leibniz Institute for Solar Physics (KIS) in Freiburg with the Leibniz Institute for Astrophysics Potsdam, the Institute for Astrophysics Göttingen, and the Max Planck Institute for Solar System Research in Göttingen as partners, and with contributions by the Instituto de Astrofísica de Canarias and the Astronomical Institute of the Academy of Sciences of the Czech Republic. This paper made use of the IAC Supercomputing facility HTCondor (http://research.cs.wisc.edu/htcondor/), partly financed by the Ministry of Economy and Competitiveness with FEDER funds, code IACA13-3E-2493.

### ORCID iDs

J. C. Trelles Arjona  https://orcid.org/0000-0001-9857-2573
M. J. Martínez González  https://orcid.org/0000-0001-5560-7502
B. Ruiz Cobo  https://orcid.org/0000-0001-9550-6749